# A New Paradigm for Edge Reconstruction in Fractional Quantum Hall States

**Ron Sabo[1,§], Itamar Gurman[1,§], Amir Rosenblatt[1], Fabien Lafont[1], Daniel Banitt[1], Jinhong Park[2], Moty Heiblum[1], Yuval Gefen[2], Vladimir Umansky[1], and Diana Mahalu[1]**

[1]*Braun Center for Submicron Research, Dept. of Condensed Matter physics, Weizmann Institute of Science, Rehovot 76100, Israel*

[2]*Department of Condensed Matter physics, Weizmann Institute of Science, Rehovot 76100, Israel*

§ *Both authors equally contributed to this work*

## ABSTRACT

**Questions on the nature of edge reconstruction and 'where does the current flow' in the quantum Hall effect (QHE) have been debated for years. Moreover, the recent observation of proliferation of 'upstream' neutral modes in the fractional QHE raised doubts about the present models of edge channels. In this article we focus on hole-conjugate states, $\nu = \frac{2}{3}$ and $\nu = \frac{3}{5}$, and present a new picture of their edge reconstruction. For example, while the present model for $\nu = \frac{2}{3}$ consists of a single *downstream* charge channel with conductance $\frac{2}{3}\frac{e^2}{h}$ and an *upstream* neutral mode, we show that the current is carried by two separate *downstream* edge channels, each with conductance $\frac{1}{3}\frac{e^2}{h}$, accompanied by upstream neutral mode(s). We find that if the two downstream channels are not equilibrated, inter-mode equilibration (via particle exchange) takes place over a distance of microns, with the two channels effectively behaving as a single channel. Moreover, the inter-channel equilibration is accompanied by an excitation of upstream neutral modes. In turn, the counter-propagating neutral modes, moving in close proximity to the charge modes, fragment into propagating charges, inducing thus downstream current fluctuations with zero net current – a novel mechanism for non-equilibrium noise. This**

**unexpected edge reconstruction underlines the need for better understanding of edge reconstruction and energy transport in all fractional QHE states.**

## Introduction

It is well accepted that transport of charge in the quantum Hall effect (QHE) is mediated by *downstream* chiral edge channels, while the bulk is *incompressible*. The Hall conductance exhibits plateaus centered at rational filling factors $v$, $G_H = v\frac{e^2}{h}$ ($e$ – electron's charge; $h$ - Planck's constant), accompanied by a vanishing longitudinal conductance (and resistance). Whereas the edge profile of particle-like (Laughlin's) fractional states was expected to mimic that of integer states [1, 2], the hole-conjugate states were believed to be more complex. For the latter, substantial 'edge-reconstruction' was expected, with added *upstream* chiral edge modes [3]. A new ground state due to unavoidable inter-channel scattering and interactions is established, with upstream neutral modes joining the downstream charge channels [4, 5, 6]. However, the recent observation of upstream chiral neutral modes in particle-like fractions [7, 8], accompanied by energy flow through the incompressible bulk, raised doubts of the validity of the presently accepted edge models.

Here, we studied two hole-conjugate states, $v = \frac{2}{3}$ (in some detail) and $v = \frac{3}{5}$. For the former, Girvin and MacDonald proposed [9, 3] an edge structure composed of two counter-propagating channels: a downstream chiral channel with conductance $\frac{e^2}{h}$ and an upstream chiral channel with conductance $-\frac{1}{3}\frac{e^2}{h}$ (the minus sign stands for counter-propagating, Fig. 1a, I). With the upstream not observed [10], Kane *et al.* [4, 5] (for short, KFP), proposed taking into account inter-channel interactions accompanied with inter-channel scattering - an equilibrated new ground state composed of a downstream charge channel (with conductance $\frac{2}{3}\frac{e^2}{h}$) and an upstream neutral mode (with zero net electric current, see Fig. 1a, II). Other proposals consisted of:(*i*) Two downstream edge channels, each with conductance $\frac{1}{3}\frac{e^2}{h}$ [2, 11]; (*ii*) Two counter-propagating edge channels added to Girvin-MacDonald's model at the sample's edge, yielding an edge composed of (in order from bulk-to-edge): $-\frac{1}{3}\frac{e^2}{h}, \frac{e^2}{h}, -\frac{1}{3}\frac{e^2}{h}, \frac{1}{3}\frac{e^2}{h}$ charge channels [12, 13]. In that model, depicted in Fig. 1a III, the inner three edge channels were expected to renormalize to a single downstream charge mode with conductance $\frac{1}{3}\frac{e^2}{h}$ accompanied by two near-by upstream neutral modes, and an additional downstream charge channel with conductance $\frac{1}{3}\frac{e^2}{h}$ near the sample's edge.

On the experimental front, recent observations of upstream neutral modes [6, 7, 14, 15, 16] supported the KFP model of the $v = \frac{2}{3}$ state [5]. However, Bid *et al.* [17], while studying partitioning of the downstream charge channel for $v = \frac{2}{3}$ in a quantum point contact (QPC), reported a $\frac{1}{3}\frac{e^2}{h}$ conductance plateau (at QPC transmission $t = \frac{1}{2}$) with strong noise. Since shot noise results from stochastic partitioning of a noiseless particle stream, the noise-on-plateau, leads to a discrepancy in our present understanding, and thus triggered our present study.

**Results**

Figures 1b & 1c portray 'two-QPC' configurations; one with $L = 9\mu m$ and the other with $L = 0.4\mu m$, which are employed in order to understand this phenomenon. These two versions were fabricated in a high mobility 2DEG embedded in high mobility GaAs-AlGaAs heterostructures. The proposed description of the chiral edge channels plotted in the figures will be justified below.

The dependence of the total transmission $t_{S1 \to D1}$ ($L = 9\ \mu m$) on the split-gate voltage $V_{QPC2}$ is plotted in Fig. 2a. Following the '$t_{S1 \to D2} = \frac{t_1}{2}$ conductance plateau', observed for all values of $t_1$ once $t_2 = \frac{1}{2}$, we find it obeys the generalized total transmission $t_{S1 \to D1} = t_1 \times t_2$. This observation supports a 'single downstream-like' charge mode behavior. Similar results were obtained for $L = 5.3\mu m$ (not shown). A striking difference is observed with the configuration $L = 0.4\ \mu m$ (Fig. 2b). Here, the 'conductance plateau' $t_{S1 \to D1} = \frac{1}{2}$ remains as long as $t_2 = \frac{1}{2}$ and $t_1 \geq \frac{1}{2}$ (with $t_{S1 \to D1} \sim t_1$ for $t_1 < \frac{1}{2}$). This is consistent with two independent, spatially separated, unequilibrated downstream charge channels; each with conductance $\frac{1}{3}\frac{e^2}{h}$.

Specifically, for $t_1 = t_2 = t_{S1 \to D1} = \frac{1}{2}$ no current arrives at $D2$ ($t_{S1 \to D2} < 10^{-3}$; see Supplementary Information). Evidently, these two outcomes suggest that equilibration, via inter-edge charge tunneling, takes place at a length scale of a few micrometers (namely less than $5.3 \mu m$).

Is the observed edge reconstruction in the $\nu = \frac{2}{3}$ state unique to this state? Testing similar 'two-QPC' configurations at $\nu = \frac{3}{5}$ resulted in qualitatively similar observations. A single QPC $\frac{1}{3}\frac{e^2}{h}$ conductance plateau, corresponding to $t = \frac{5}{9}$, was observed (blue line in Figs. 2c & 2d); also laden with noise. Very much like in $\nu = \frac{2}{3}$, the 'two-QPC' configuration with $L = 0.4 \ \mu m$ proved the presence of unequilibrated channels (Fig. 2d); while in the $L = 5.3 \ \mu m$ device, full equilibration took place (Fig. 2c). In identical measurements at $\nu = 2$ (two integer edge channels), and at $\nu = 2/5$ (two composite fermion edge channels) [18], the edge channels remained unequilibrated even at $L = 9 \ \mu m$ (see Supplementary Section S1).

To further probe the inter-edge tunneling, we fabricated a small electronic Fabry-Perot interferometer (FPI), with an area $400 \times 400$ nm$^2$ defined by two QPCs and modified by charging the *plunger gate* with $V_P$ (Fig. 3a). Interference of such small FPI is known to be dominated by Coulomb interactions [19, 20, 21, 22].

Operating at filling $\nu = \frac{2}{3}$, with the two QPCs strongly pinched ($t_{QPC1}, t_{QPC2} \ll 1$), ubiquitous periodic conductance peaks were observed as function $V_P$ (Fig. 3b). These peaks are known not to show magnetic field ($B$) dependence if the 'interfering' channel is the outer most one (see Supplementary Section S2) [19]. Adhering to the two-channel model, we tuned each QPC to $t_{QPCi} = \frac{1}{2}$ (with the two '1/3 channels' sketched in Fig. 3a). Evidently, the total transmission $t_{FPI} = \frac{1}{2} (G_{FPI} = \frac{1}{3}\frac{e^2}{h})$, with the outer '1/3 channel' being fully transmitted. Scanning $V_P$ revealed a series of strong quasi-periodic conductance *dips* protruding down from $G_{FPI} = \frac{1}{3}\frac{e^2}{h}$ (Fig. 3c). We account these dips to resonant tunneling. While the outer-most edge channel (red line at bottom and blue line at top, Fig. 3a) is allowed to pass freely through the FPI, the inner channel (white line) is fully confined, and thus quantized. A very small, but finite, tunneling probability between the outer and inner channels allows resonant tunneling from the bottom (red) channel to the upper (blue) channel via the inner (white) channel. In other words, every time a quantized state in the confined inner edge channel (white) is aligned with the Fermi energy of the outer edge channel, quasiparticles will backscatter via resonant tunneling and arrive at $D2$ instead of $D1$. Significant conductance dips will be evident at sufficiently low temperatures.

A 2D $V_P - B$ plot of the conductance dips exhibits the familiar Coulomb dominated behavior of the 'inner dot' (Fig. 3d) [21, 22]. The extracted area from the $B$ periodicity, $A = \frac{\phi_0}{\Delta B} = 0.11\ \mu m^2$, is in fair agreement with the lithographic area. Recalling the $B$ independent transmission for the strongly partitioned outer-most channel (not shown here), the 'two-channel' model for the $\nu = \frac{2}{3}$ is reconfirmed. The temperature dependence of both dips (Fig. 3c) and peaks (Fig. 3b) qualitatively endorses tunneling of fractionally charged quasi particles, rather than electrons, between the two edge channels (see Supplementary Section S3).

Although the above results conclusively support the formation of two charge modes, thus challenging the broadly accepted KFP picture, they still do not provide an explanation for the 'noise on the plateau' reported first by Bid *et al.* [17]. We performed noise measurements with the 'two-QPC configuration' at $L = 0.4\ \mu m$ (Fig. 1c). Once the two QPCs are tuned to their conductance plateau ($t_1 = t_2 = \frac{1}{2}$), current from $S1$ (red lines) did not arrive at D2; however, significant current fluctuations were measured in $D2$ (red crosses in Fig. 4a). For comparison, we show the measured excess noise in $D2$ (blue dots) when $t_1 = 1$ and $t_2 = \frac{1}{2}$ (effectively, a single QPC configuration). The two noise plots, one without net current and the other with, are surprisingly similar.

Similar noise measurements had been repeated for $\nu = 3/5$ state, with the two QPCs tuned to the conductance plateau $\frac{1}{3}\frac{e^2}{h}$ ($t_1 = t_2 = \frac{5}{9}$). Again, strong current fluctuations, with null net current, were observed (see Supplementary Section S6). Noting, that such current fluctuations are never observed in conductance plateaus of integer or a particle-like fractional states.

In general, stochastic partitioning of an impinging 'quiet current' at a QPC leads to shot noise with spectral density $S = 2qIt(1-t)\alpha(T)$, where $q$ is the partitioned charge, $I$ the impinging DC current, $t$ the transmission of the partitioned channel, and $\alpha(T)$ a temperature dependent reduction factor [23]. It is convenient to define a *Fano factor* $F = \frac{S}{2eIt(1-t)\alpha(T)}$, being effectively the partitioned charge, $q/e$ [24, 25, 26]. When the current fluctuations at the $t = \frac{1}{2}$ plateau at the $\nu = \frac{2}{3}$ state were analyzed, assuming a single charge channel with conductance $\frac{2}{3}\frac{e^2}{h}$, yielded $F = \frac{2}{3}$ at $T \sim 20$ mK and $F = \frac{1}{3}$ at higher temperatures ($T \sim 100$mK) [17]. Evidently, this analysis is irrelevant with the presently understood edge reconstruction. Yet, it is convenient to define an 'effective Fano factor', by replacing the $t(1-t)$ term with a constant $\frac{1}{4}$ and considering $I$ to be the total current leaving the source; namely, $F_{eff} = \frac{2S}{eI\alpha(T)}$. Note, that while for a single QPC $F_{eff} = \frac{2}{3}$, in the 'two-QPC' configuration, with $t_1 = t_2 = \frac{1}{2}$ (red crosses, Fig 4a), $F_{eff} \sim \frac{1}{2}$.

Sourcing from $S3$ with $t_1 = t_2 = \frac{1}{2}$, led to even more puzzling results. Current carried by the outer (inner) edge channel reached only $D2$ ($D1$) - with no current reaching $D3$. Yet, substantial current fluctuations were measured in $D3$ with $F_{eff} = 0.40$. Evidently, this currentless - noise must result from upstream neutral mode(s) [14]. Noise measurements in all nine possible configurations, with $t_1 = t_2 = \frac{1}{2}$, are summarized in the Fig. 6, and compared below to predictions of a simplified theoretical model.

**Discussion**

Inspired by the correlation between the noise-on-plateau and the evident presence of equilibration process between the two 1/3 - modes, we propose a new mechanism for the generation of the non-equilibrium noise, proportional to the injected current; due to interplay between counter propagating charge and neutral modes. Start with a 'hot' (noiseless) current impinging from $S1$ on a QPC tuned to the $t = \frac{1}{2}$ plateau (red lines, Fig. 4b), with a 'cold' current arriving from the grounded $S2$ (blue lines). Separating the inner and the outer channels at the QPC leads to two co-propagating pairs; 'hot' and 'cold' (moving towards $D1$ and $D2$). Since the propagation distance of each pair is tens of micrometers long, equilibration within each pair takes place (purple lines). This process excites upstream neutral modes (dotted lines, yellow), which counter-propagate towards the QPC. These counter propagating modes fragment into particle-hole pairs in the propagating charge channels. A similar, yet more complicated scenario, with the 'two-QPC' setup, with current emanating from $S1$, is shown in Fig. 4c. The remaining question is the observed quantized value of $F_{eff}$.

We now introduce a quantitative model that accounts for the quantized value of $F_{eff}$ in the single-QPC setup at $t = \frac{1}{2}$ plateau (Fig. 5). Generalization of this model that addresses the different Fano factors in the 'two-QPC' geometry is provided in Supplementary Section S7. The noise-generating mechanism consists of a two-step process (we refer to this as a *first hierarchy process*): (*i*) Charge equilibration accompanied by a generation of neutral mode excitations ('neutralons' or 'anti-neutralons', see Supplementary Section S8); (*ii*) Fragmentation (*i.e.*, decay) of neutralons, leading to a stochastic creation of quasi-particle / quasi-hole pairs in the charged channels.

The edge of the $v = 2/3$ fraction is now expected to support four chiral channels [13]: two downstream charged modes, each with conductance $\frac{1}{3}\frac{e^2}{h}$, and two inner upstream neutral modes (denoted as yellow lines in Fig. 5), supporting neutralons (and anti-neutralons). Let us assume that during time $\tau$, $S1$ emits $2N$ quasi-particles ($N$ in each channel), each of charge $e/3$, giving rise to an emitted current $I = \frac{2Ne}{3\tau}$. Having the QPC set to $t = \frac{1}{2}$ plateau, one channel is fully reflected while the other is fully transmitted. Transmitted (reflected) through (off) the QPC, 'hot' channels (denoted by red solid lines in Fig. 5) are moving in parallel to 'cold' channels emanating from grounded contacts (denoted as blue solid lines). Equilibration between the two channels takes place in both outgoing trajectories (Fig. 5a). Assuming the latter are sufficiently long, complete equilibration implies that $N/2$ quasi-particles tunnel from the 'hot' channel to the 'cold' one. Close examination of the tunneling operators reveals that each such tunneling event is accompanied with the generation of two neutralons (Supplementary Information). Being inner upstream channels, they subsequently fully reflect by the QPC (Fig. 5b). Eventually the neutralons decay through a process that converts a pair of them into a quasi-particle / quasi-hole (quasi-hole / quasi-particle) in the respective (parallel flowing) charge modes (Fig. 5c).

These stochastic processes occur with equal probabilities. The newly added quasi-particles and quasi-holes flow towards the QPC where they split - propagating thereafter towards different drains (Fig. 5d). Thus the average current in each charge channel is unchanged by the decay of the neutralons, yet these stochastic processes generate non-equilibrium noise. This noise is proportional to the injected current; hence *shot-noise*-like.

To characterize this noise quantitatively we introduce $N/2$ random variables $a_i$ ($b_i$) $i = 1, \ldots, N/2$, referring to the neutralon decay on the lower left (upper right) corner of Fig. 5c. The variables assume the values +1 or -1, each with probability $\frac{1}{2}$, where +1 represents a creation of a quasi-particle in the outer channel and a quasi-hole in the inner channel; while -1 represents the opposite process. The total charge $Q_{D1}$ arriving at $D1$ during the time interval $\tau$ is given by $Q_{D1} = \frac{e}{3}\left(N + \sum_{i=1}^{N/2} a_i - \sum_{i=1}^{N/2} b_i\right)$, with the average charge $\overline{Q_{D1}} = \frac{e}{3}N$, and its variance is $\overline{(\delta Q_{D1})^2} = \overline{(Q_{D1} - \overline{Q_{D1}})^2} = \frac{e^2 N}{9}$. Here, we assume that all of the neutralons decay into quasi-particles and quasi-holes. The 'zero frequency' excess noise is $S_{D1} = 2 \lim_{\tau \to \infty} \frac{\overline{(\delta Q_{D1})^2}}{\tau}$, resulting in $F_{eff} = \frac{2}{3}$, in agreement with the experimental value. The same magnitude of noise is measured (and calculated) at $D2$.

Although this model explains well the quantized value of the noise in a single QPC, when it comes to the 'two-QPC' setup there are discrepancies between our theory (cf. $F_{th}^{(1)}$ in the table at Fig. 6) and the experimentally observed $F_{eff}$'s (cf. $F_{exp}$). These discrepancies may be resolved when *second hierarchy processes* are included. Consider, for example, the configuration depicted in Fig 4c, featuring $S1$ as the biased source (all other contacts are grounded). First hierarchy processes result in a noiseless inner channel and a noisy outer channel (both at same chemical potential) flowing into $D2$. Yet, local fluctuations in the charge density of the two channels lead to further equilibration processes (facilitating quasi-particle tunneling from a channel whose instantaneous density, at a given point, is higher than that of the other channel). Such tunneling processes generate additional neutralons (or anti-neutralons), which flow upstream from $D2$ to $S2$. These neutralons annihilate into quasi-particle / quasi-hole pairs at the charge channels, making further contributions to the noise in drains $D2$ and $D3$. In a simplified point-of-view, this process can be understood as equilibrating two channels at different temperatures (as opposed to the first hierarchy process, where the difference was in chemical potential). We refer to this additional two-step process, (*i*) Charge equilibration and (*ii*) Decay of neutralons, as a *second hierarchy*. The consequently revised Fano factors ($F_{th}^{(2)}$) are displayed in the table (Fig. 6). The second hierarchy process indeed improves the agreement between experiment and theory. For details of the second hierarchy see Supplementary Information.

**Summary**

We portrayed here a thorough experimental and theoretical study of transport in fractional hole-conjugate states. Specifically, in the $\nu = 2/3$ case, we proved that the current is carried by two spatially separated, co-propagating, downstream edge channels (each with conductance $\frac{1}{3}\frac{e^2}{h}$). Moreover, when these edge channels are out of equilibrium, they equilibrate due to inter-edge tunneling of fractionally charged quasi-particles, with a typical equilibration length of a few micrometers. Further, we observed unexpected shot noise like fluctuations, which we interpret as a unique interplay between charge and neutral modes. Namely, equilibration of charge modes excites neutral modes, which in turn decay and induce noise in the charge modes. In addition, our proposed theoretical model provides a quantitative estimate of the low temperature Fano factor, which agrees well with the experimental results in the 'single-QPC' setup, and leads to partial agreement with the noise measured in the 'two-QPC' setup.

These results suggest a new paradigm with a new approach to looking at fractional states and the behavior of edge modes, especially in cases where reconstruction at the edge takes place leading to formation of counter-propagating modes [7]. The yet unresolved discrepancies mentioned in the paper, temperature dependent Fano factor or inconsistencies in the 'two-QPC' configuration, call for more theoretical and experimental investigations, which will continue to pave the road to a more complete understating of the FQHE.

## Methods:

**Sample fabrication:** The samples were fabricated in a GaAs-AlGaAs heterostructures, embedding a 2DEG, with areal density of $(1.2 - 2.5) \times 10^{11}$ cm$^{-2}$ and 4.2 K 'dark' mobility $(3.9 - 5.1) \times 10^6$ cm$^2$V$^{-1}$s$^{-1}$, $(70 - 116)$ nm below the surface. The different gates were defined with electron beam lithography followed by deposition of Ti/Au. Ohmic contacts are made from annealed Au/Ge/Ni. The sample was cooled to $20 mK$ (40 mK for the $\nu = \frac{3}{5}$ data) in a dilution refrigerator.

**Measurement technique**: Conductance measurements were done by applying an AC signal with an $\sim 1 \, \mu V_{RMS}$ excitation at 700 KHz in the relevant source, which resulted in drain voltage $V_D = I_D R_H$, with $R_H$, the Hall resistance. The drain voltage was filtered using an LC resonant circuit and amplified by homemade voltage preamplifier (cooled to 1K) followed by a room temperature amplifier (NF SA-220F5).

## Author Contributions:

R.S, I.G, A.R and M.H designed the experiment, R.S, I.G, A.R, F.L, D.B and M.H preformed the measurements, R.S, I.G, A.R, F.L, and M.H did the analysis, J.P and Y.G developed the theoretical model. R.S, I.G, A.R, F.L, J.P, M.H and Y.G wrote the paper. V.U. grow the 2DEG and D.M. was responsible on the e-beam lithography.

## Acknowledgements:

M.H. acknowledges the partial support of the Minerva foundation, grant no. 711752, the German Israeli Foundation (GIF), grant no. I-1241- 303.10/2014, and the European Research Council under the European Community's Seventh Framework Program (FP7/2007-2013)/ERC Grant agreement No. 339070. Y.G. acknowledges the partial support of DFG Grant No. RO 2247/8-1, the Minerva foundation, the Russia-Israel IMOS project and by CRC183 of the DFG. I.G. is grateful to the Azrieli Foundation for the award of an Azrieli Fellowship.

**References**:

[1]   X.-G. Wen, "Theory of the edge states in fractinal quantum Hall effects," *Int. J. Mod. Phys. B,* vol. 06, p. 1711, 1992.

[2]   C. W. J. Beenakker, "Edge channels for the fractional quantum Hall effect," *Phys. Rev. Lett.,* vol. 64, p. 216, 1990.

[3]   A. H. MacDonald, "Edge states in the fractional quantum Hall effect regime," *Phys. Rev. Lett. ,* vol. 64, p. 220, 1990.

[4]   C. L. Kane, M. P. A. Fisher and J. Polchinski, "Randomness at the Edge: Theory of Quantum Hall Transport at Filling v=2/3," *Phys. Rev. Lett. ,* vol. 72, pp. 4129-4132 , 1994.

[5]   C. L. Kane and M. P. A. Fisher, "Impurity scattering and transport of fractional quantum Hall edge states," *Phys. Rev. B ,* vol. 51, p. 13449 , 1995.

[6]   A. Bid, N. Ofek, H. Inoue, M. Heiblum, C. Kane, V. Umansky and D. Mahalu, "Observation of Neutral Modes in the Fractional Quantum Hall Regime.," *Nature ,* vol. 466, p. 585 , 2010.

[7]   H. Inoue, A. Grivnin, Y. Ronen, M. Heiblum, V. Umansky and D. Mahalu, "Proliferation of neutral modes in fractional quantum Hall states," *Nat. Comm.,* vol. 5, p. 4067, 2014.

[8]   C. Altimiras, H. le Sueur, U. Gennser, A. Anothore, A. Cavanna, D. Mailly and F. Pierre, "Chargeless heat transport in the fractional quantum Hall regime," *Phys. Rev. Lett.,* vol. 109, p. 026803 , 2012.

[9]   S. M. Grivin, "Particle-hole symmetry in the anomalous quantum Hall effect," *Phys. Rev. B,* vol. 29, p. 6012(R), 1984.

[10] R. C. Ashoori, H. L. Stormer, L. N. Pfeiffer, W. Baldwin and K. West, "Edge magnetoplasmons in the time domain," *Phys. Rev. B,* vol. 45, p. 3894, 1992.

[11] A. M. Chang, "A unified transport theory for the integral and fractional quantum hall effects: Phase boundaries, edge currents, and transmission/reflection probabilities," *Solid State Communications,* vol. 74, no. 9, pp. 871-876, 1990.

[12] Y. Meir , "Composite edge states in the v=2/3 fractional quantum Hall regime," *Phys. Rev. Lett.,* vol. 72, p. 2624, 1994.

[13] J. Wang, Y. Meir and Y. Gefen, "Edge Reconstruction in the v=2/3 Fractional Quantum Hall State," *Phys. Rev. Lett. ,* vol. 111, p. 246803, 2013.

[14] Y. Gross, M. Dolev, M. Heiblum, V. Umansky and D. Mahalu, "Upstream neutral modes in the fractional quantum Hall effect regime: heat waves or coherent dipoles?," *Phys. Rev. Lett.,* vol. 108, p. 226801 , 2012.

[15] I. Gurman, R. Sabo, M. Heiblum, V. Umansky and D. Mahalu, "Extracting net current from an upstream neutral mode in the fractional quantum Hall regime," *Nat. Comm.,* vol. 3, p. 1289, 2012.

[16] V. Venkatachalamy, S. Hart, L. Pfeiffer, K. West and A. Yacoby, "Local Thermometry of Neutral Modes on the Quantum Hall Edge," *Nature Phyiscs,* vol. 8, p. 676, 2012.

[17] A. Bid, N. Ofek, M. Heiblum, U. Umansky and D. Mahalu, "Shot Noise and Charge at the 2/3 Composite Fractional Quantum Hall State," *Phys. Rev. Lett.,* vol. 103, p. 236802, 2009.

[18] J. K. Jain, Composite Fermins, Cambridge University Press, 2007.

[19] N. Ofek, A. Bid, M. Heiblum, A. Stern, V. Umansky and D. Mahalu, "The Role of Interactions in an Electronic Fabry-Perot Interferometer Operating in the Quantum all Effect Regime," *PNAS ,* vol. 107, p. 5276, 2010.

[20] D. T. McClure , W. Chang, C. M. Marcus, L. N. Pfeiffer and K. W. West, "Fabry-Perot Interferometry with Fractional Charges," *Phys. Rev. Lett.,* vol. 108, p. 256804, 2012.

[21] B. Rosenow and B. I. Halperin, "Influence of Interactions on Flux and Back-Gate Period of Quantum Hall Interferometers," *Phys. Rev. Lett.,* vol. 98, p. 106801, 2007.

[22] B. I. Halperin, A. Stern, I. Neder and B. Rosenow, "Theory of the Fabry-Pérot quantum Hall interferometer," *Phys. Rev. B,* vol. 83, p. 155440, 2011.

[23] T. Martin and R. Landauer, "Wave-packet approach to noise in multichannel mesoscopic systems," *Phys. Rev. B,* vol. 45, p. 1742, 1992.

[24] R. de-Picciotto, M. Reznikov, M. Heiblum, V. Umansky, G. Bunin and D. Mahalu, "Direct observation of a fractional charge," *Nature,* vol. 389, p. 162, 1997.

[25] L. Saminadayar, D. C. Glattli, Y. Jin and B. Etienne, "Observation of the e/3 Fractionally Charged Laughlin Quasiparticle," *Phys. Rev. Lett.,* vol. 79, p. 2526, 1997.

[26] M. Dolev, M. Heiblum, V. Umansky, A. Stern and D. Mahalu, "Observation of a quarter of an electron charge at the : ν = 5/2 quantum Hall state," *Nature,* vol. 452, p. 829, 2008.

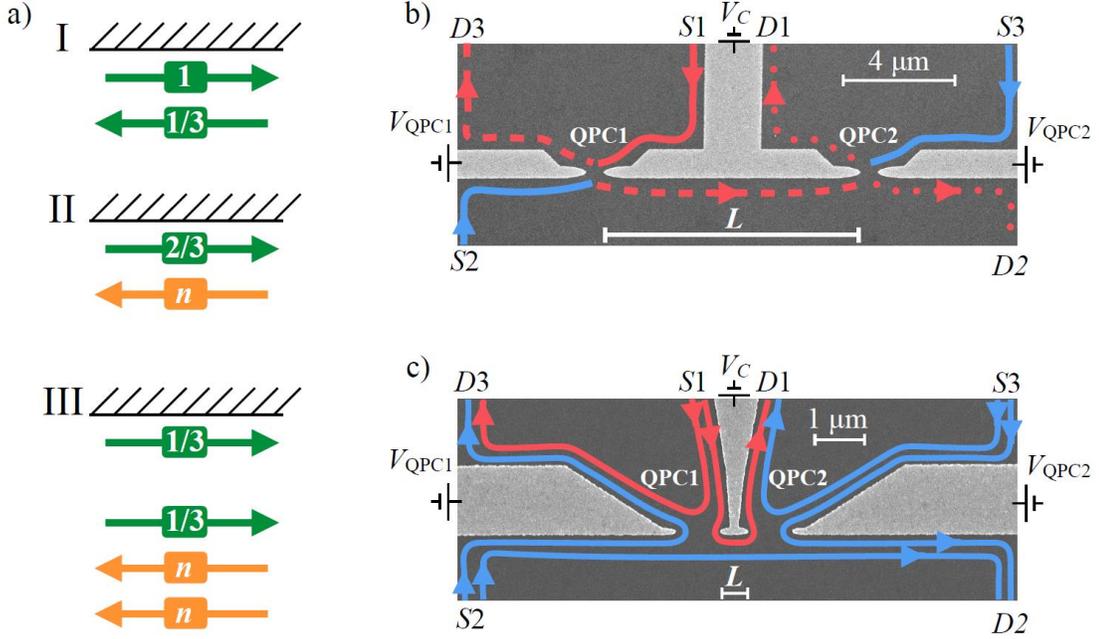

**Figure 1: the different theoretical modes of $\nu = 2/3$ and the observation of two separate charge modes and the way to differentiate between one and two channels**

a) Three different theoretical descriptions of the $\nu = \frac{2}{3}$ edge structure. I: an upstream charge mode with conductance $\frac{1}{3}\frac{e^2}{h}$ followed with a downstream charge mode with conductance $\frac{e^2}{h}$ (*MacDonald picture*). II: mixing of the two above mentioned channels, due to impurity scattering, resulting with a single downstream charge mode and an upstream neutral mode (*KFP picture*). III: four-edge modes model is obtained by adding two counter-propagating $\frac{1}{3}\frac{e^2}{h}$ edge modes to I, and then introducing scattering (*Meir picture*).

b) A SEM image of the two consecutive QPCs configuration. Current impinging from $S1$ (clockwise chirality) can be partitioned at both QPCs, which are controlled by the voltages $V_{QPC1}$, $V_{QPC2}$ and $V_C$ (the common middle gate), and arrive at either one of the three drains in the system ($D1 - D3$). Similarly, one can source from $S2$ or $S3$ and measure the outcome in the various drains. In the case of a single charge mode, the total transmission from $S1$ to $D1$ (red edge) is the product of the two transmission processes, specifically once both are tuned to half transmission the total outcome will be one quarter. In addition, the transmission to $D2$ will also be one quarter.

c) Similar to (b), yet with two charge modes. In this case the resulting transmission from $S1$ to $D1$ is highly dependent on which channel each QPC is tuned to. Specifically, if both QPCs are tuned to half transmission (as illustrated here) it means the two channels are split at each QPC and the overall result will be one half. Furthermore, in this case no net current will arrive at $D2$.

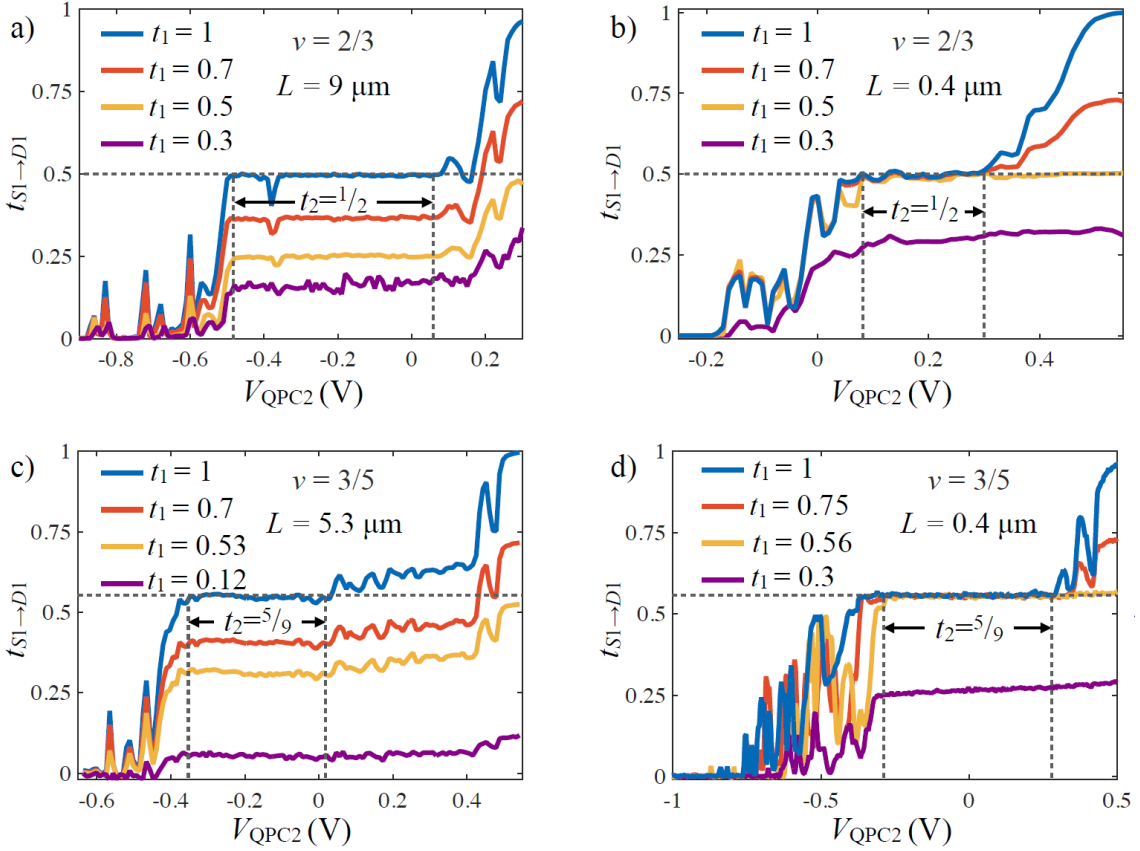

**Figure 2: Comparison between the $L = 9\ \mu m$ and the $L = 0.4 \mu m$ devices:**

a) Transmission from S1 to D1 ($t_{S1\rightarrow D1}$) as function of $V_{QPC2}$ for different $t_1$'s for the $L = 9 \mu m$ device at $\nu = \frac{2}{3}$. The clear plateau at $\frac{t_1}{2}$ is a signature of the successive partitioning of a single edge mode. Dashed vertical lines mark the region where QPC2 exhibits a plateau.

b) Same as (a) for the $L = 0.4\ \mu m$ device. The value of the plateau is $\frac{1}{2}$ as long as $t_1 > \frac{1}{2}$; a signature of transport through two independent edge modes.

c) $t_{S1->D1}$ as function of $V_{QPC2}$ for different $t_1$'s for a $L = 5.3\ \mu m$ device at $\nu = \frac{3}{5}$. The $t_1 = 1$ case (blue curve) shows a transmission plateau of $\frac{5}{9}$, which corresponds to an outer channel with $\frac{1}{3}\frac{e^2}{h}$ conductance. Plateaus at different value (proportional to $t_1$) proves the full equilibration between the two modes. Black dashed line shows the plateau value of $\frac{5}{9}$.

d) Same as (c) for the $L = 0.4 \mu m$ device. The value of the plateau is $\frac{5}{9}$ as long as $t_1 \geq \frac{5}{9}$, proving that also in $\nu = \frac{3}{5}$ the edge is made out of two independent edge modes.

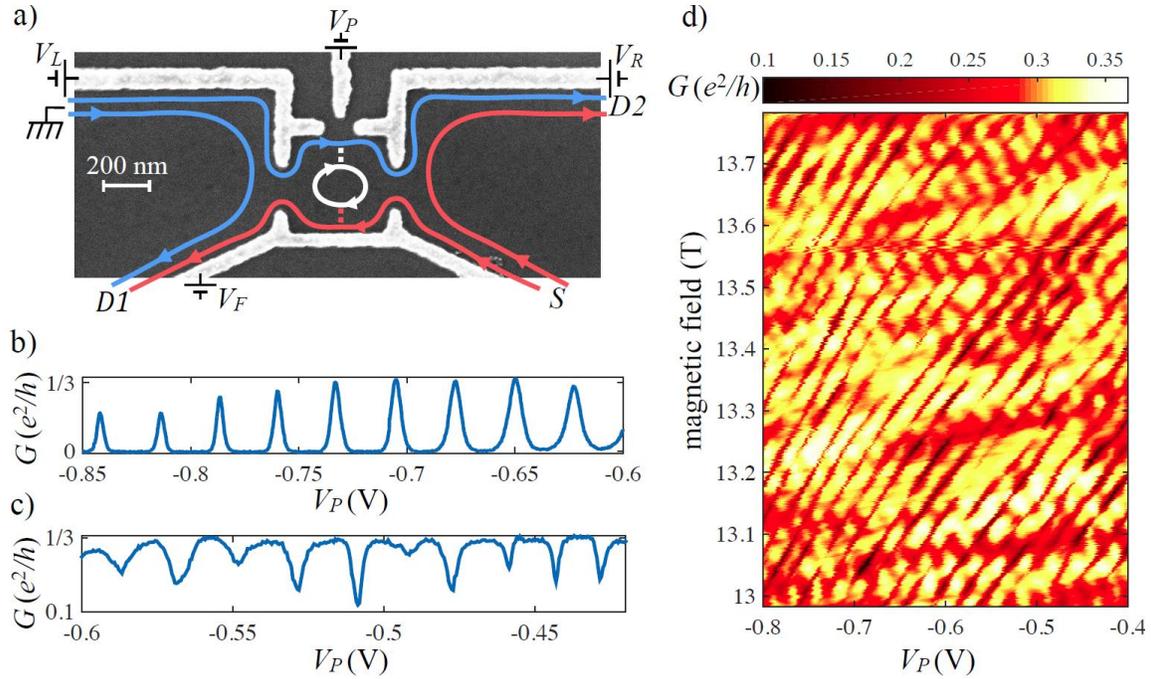

**Figure 3: Fabry-Perot interferometer (FPI) geometry:**
  a) SEM image of the device together with the edge propagation scheme for the case of $t_{QPC1} = t_{QPC2} = \frac{1}{2}$. The outer edge mode, either biased (red) or unbiased (blue), flows unperturbed through both QPCs that define the FPI. The inner edge mode is fully reflected at the QPCs, forming an island inside the FPI (white). Dotted lines denote the tunneling between the two modes.
  b) Series of conductance (Coulomb blocked) peaks observed for the case of $t_{QPC1}, t_{QPC2} \ll 1$.
  c) Series of quasi-periodic dips below $t_{FPI} = \frac{1}{2}$ for the case of $t_{S \to D1}$. These dips are the result of the resonant tunneling from the lower outer biased channel into the FPI and to the upper unbiased outer channel .
  d) Conductance dips evolution in magnetic field. Equal phase lines have anti Aharonov-Bohm behavior as expected from interference of an inner edge mode. The area extracted from the magnetic field periodicity fits the area of the interferometer.

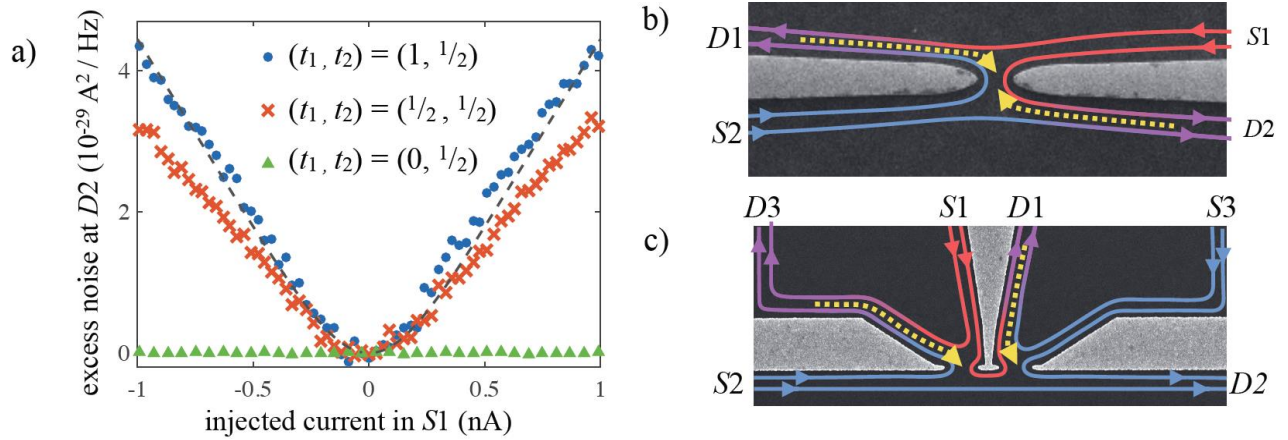

**Figure 4: Finite current fluctuations accompanying null net current, and a schematic model for the interplay between charge and neutral modes**

a) Current fluctuations measured at $D2$ as function of current injected at $S1$ when the QPCs are set to $(t_1, t_2) = \left(\frac{1}{2}, \frac{1}{2}\right)$ [red] and $\left(1, \frac{1}{2}\right)$ [blue]. In the latter case, the observed current fluctuations mimic the shot noise predicted when partitioning single charge channel with charge $e^* = \frac{2}{3}$ at half transmission and temperature of 20mK (black dashed line), thus obtaining $F = F_{\text{eff}} = 2/3$ (see text). For the $(t_1, t_2) = \left(\frac{1}{2}, \frac{1}{2}\right)$ configuration (red) we find $F_{\text{eff}} = 0.5 \pm 0.05$. As a sanity check, we show null noise once $(t_1, t_2) = \left(0, \frac{1}{2}\right)$ [green], hence QPC1 is completely closed.

b) Current injected from $S1$ (red) reached the QPC set to its $t = \frac{1}{2}$ plateau together with charge mode at ground potential (blue) emanating from $S2$. Away from the QPC, two pairs of high and low chemical potential edge channels co-propagate and thus mix (turn purple). This process is accompanied by creation of neutral excitations (yellow), which propagates back upstream towards the QPC.

c) Same as b) but for the two-QPC configuration. Neutral modes are excited only on the way to $D1$ and $D3$ flowing back towards the QPCs.

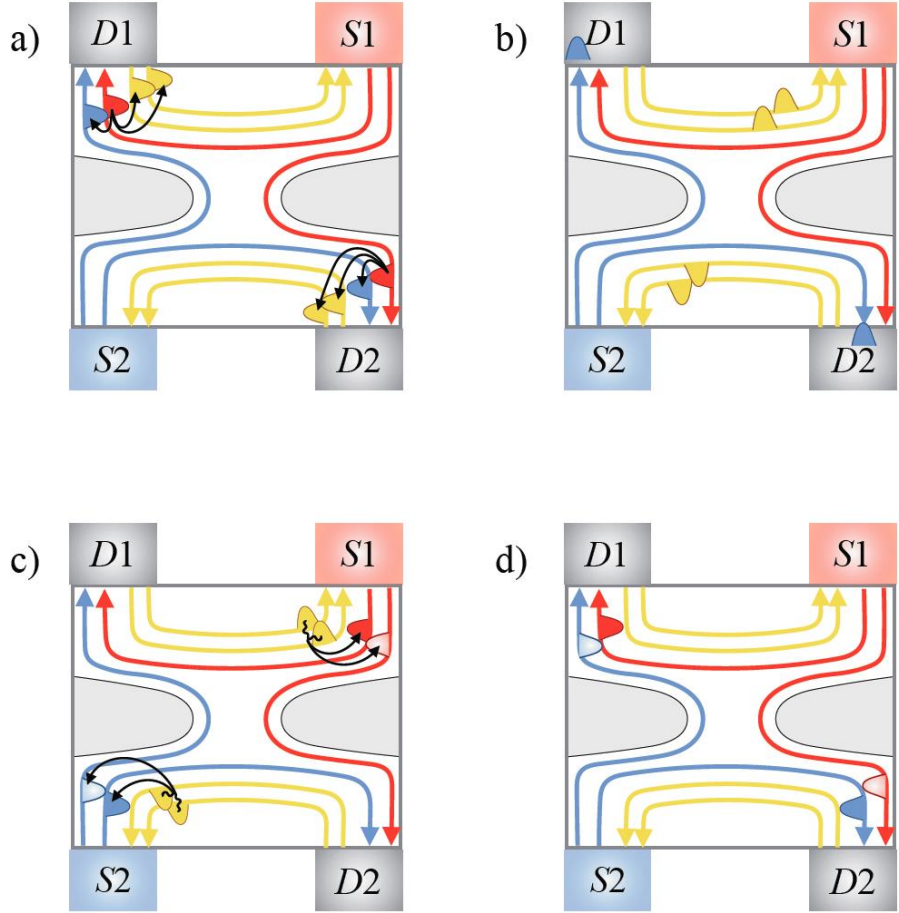

**Figure 5: Neutralon induced noise.**
A system tuned to a bulk filling factor $\nu = 2/3$ with a single QPC. The edge consists of four channels - two accommodating downstream charged modes (outer and inner), and two inner upstream neutral modes. The charge channels support quasi-particles (qp) and quasi-holes (qh), possessing a charge of $\pm e/3$. Similarly, the two neutral channels support neutral excitations, dubbed 'neutralons' and denoted as $n_1$ and $n_2$. Biased charge channel (emanating from the source $S1$) are marked in red, unbiased charge channels are in blue and upstream neutral channels in yellow.
  a) Equilibration process involving charge tunneling and the creation of neutralons, for example $qp_{outer} \rightarrow qp_{inner} + n_1 + n_2$ (lower right corner) or $qp_{inner} \rightarrow qp_{outer} + n_1 + n_2$ (upper left corner). A list of all equilibration processes is presented in the Supplemental Information.
  b) The tunneling qp's flow downstream to either drains, while the two excited neutrons flow upstream and fully reflect from the QPC, back towards $S1$ and $S2$.
  c) Each neutralon pair decays to give rise to a qp/qh pair, which is randomly split between the two charge modes.
  d) The excited charge modes flow back to the QPC, the inner one fully reflects and the outer fully transmits, not influencing the average currents at the drains yet contributing to the measured noise (current fluctuations).

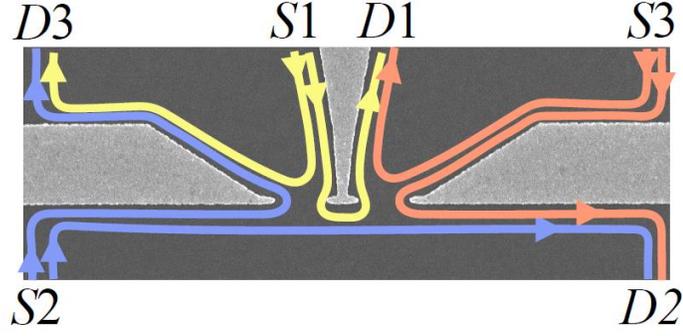

| | S1 | | | S2 | | | S3 | | |
|---|---|---|---|---|---|---|---|---|---|---|
| | $t$ | $F_{\text{exp}}$ | $F_{\text{th}}^{(1)}$ | $F_{\text{th}}^{(2)}$ | $t$ | $F_{\text{exp}}$ | $F_{\text{th}}^{(1)}$ | $F_{\text{th}}^{(2)}$ | $t$ | $F_{\text{exp}}$ | $F_{\text{th}}^{(1)}$ | $F_{\text{th}}^{(2)}$ |
| D1 | 0.5 | 0.76 | $2/3$ | $2/3$ | 0 | 0.54 | $1/3$ | $1/2$ | 0.5 | 0.51 | $1/3$ | $1/2$ |
| D2 | 0 | 0.48 | $1/3$ | $1/2$ | 0.5 | 0.50 | $1/3$ | $1/2$ | 0.5 | 0.66 | $2/3$ | $2/3$ |
| D3 | 0.5 | 0.49 | $1/3$ | $1/2$ | 0.5 | 0.68 | $2/3$ | $2/3$ | 0 | 0.40 | $1/3$ | $1/2$ |

**Figure 6: Effective Fano factors in a two-QPCs configuration**

Summary of the measured effective Fano factor ($F_{\text{exp}}$) and theoretically calculated ones evaluated based on first hierarchy processes ($F_{\text{th}}^{(1)}$) and second hierarchy processes ($F_{\text{th}}^{(2)}$), in the two-QPC configuration. For convenience, we color the channels emitted from each source (and their corresponding columns in the table) in different colors – yellow ($S1$), blue ($S2$) and red ($S3$). In each column we present the relative current arriving at each one of the drains (marked as $t$) and the various effective Fano factors. Notice the accuracy of $F_{\text{exp}}$ is $\pm 0.05$.

# Supplementary Information

## A New Paradigm for Edge Reconstruction in Fractional Quantum Hall States

Ron Sabo[1,§], Itamar Gurman[1,§], Amir Rosenblatt[1], Fabien Lafont[1], Daniel Banitt[1], Jinhong Park[2], Moty Heiblum[1], Yuval Gefen[2], Vladimir Umansky[1], and Diana Mahalu[1]

[1]*Braun Center for Submicron Research, Dept. of Condensed Matter physics, Weizmann Institute of Science, Rehovot 76100, Israel*

[2]*Department of Condensed Matter physics, Weizmann Institute of Science, Rehovot 76100, Israel*

*§ Both authors equally contributed to this work*

## S1. Edge structure in $\nu = \frac{2}{5}$

The $\nu = \frac{2}{5}$ FQH state is the $p = 2$ in the composite Fermions series $\nu = \frac{p}{2p+1}$, as such its edge structure is made out of two charge edge modes; an outer one with $G_{outer} = \frac{1}{3}\frac{e^2}{h}$ and an inner one with $G_{inner} = \frac{1}{15}\frac{e^2}{h}$ such that $G_{total} = G_{outer} + G_{inner} = \frac{2}{5}\frac{e^2}{h}$. We have repeated the measurements described in the main text using the device in Fig. 1 (two QPCs with $L = 9\mu m$ separation between them) at $\nu = \frac{2}{5}$ in order to test its edge structure. Fig S1.1 shows the transmission through a single QPC as function as the voltage applied on it. A plateau at $t = \frac{G_{outer}}{G_{total}} = \frac{5}{6}$ is clearly visible, confirming the above mentioned edge structure. Unlike the $\nu = \frac{2}{3}$ case, no current fluctuations were measured on the $t = \frac{5}{6}$ plateau.

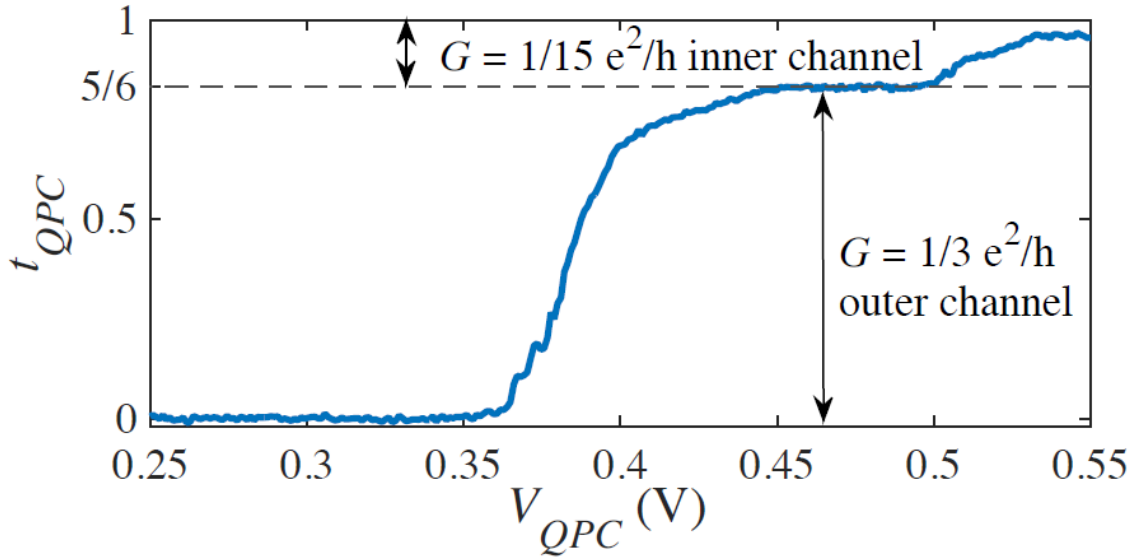

Fig S1.1: Single QPC transmission as function of voltage applied on it. A clear conductance plateau at $t = \frac{5}{6}$ ($G = \frac{1}{3}\frac{e^2}{h}$) separates the two edge channels.

Turning to the two QPC setup and setting both QPCs to their plateau value ($t_{QPC1} = t_{QPC2} = \frac{5}{6}$), no tunneling between the two edges states were observed (see Fig S1.2). In addition, no excess noise was measured in D2 in response to current injected in S1.

This result proves that the edge structure reported in the main text is unique to hole-conjugate FQH states.

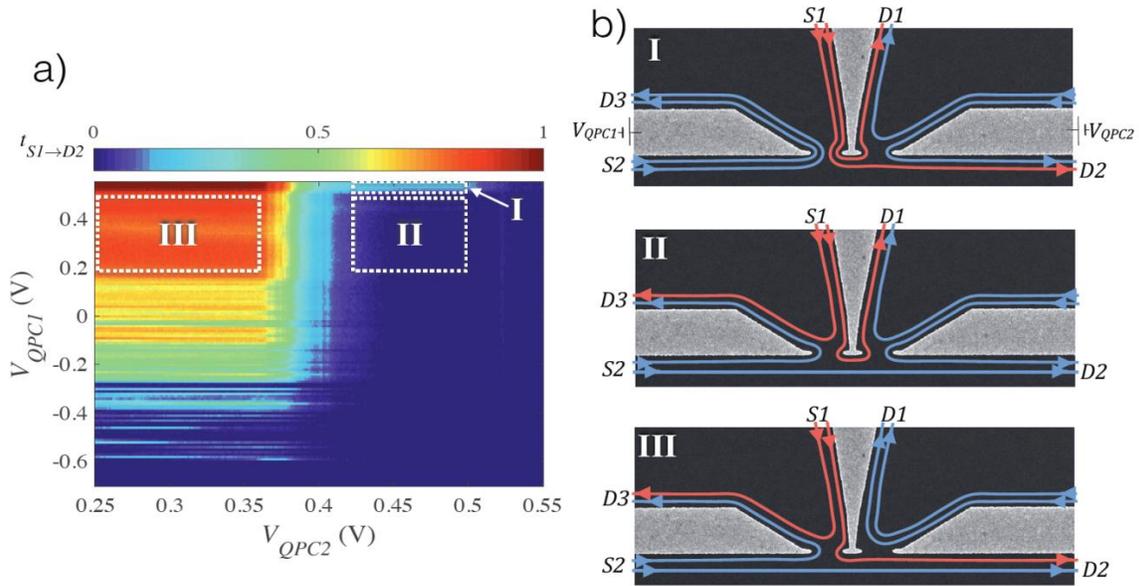

**Fig S1.2**: a) $t_{S1 \to D2}$ as function of voltage applied on both QPCs. The zero current measured when both QPC are set to their plateau value (region $II$) proves there is no tunneling between the two channels along the $9\mu m$ of co-prorogation. b) Schematic sketch of the two edge channels prorogation scheme in different areas in a) (marked $I - III$)

**S2. FPI and Coulomb Dominated Peaks**

Fig. 3a in the main text shows an SEM image of the Fabry–Pérot interferometer used also to obtain the data of this section. As noted in the main text, when both QPCs are almost completely pinched such that $t_i \ll 1$, standard periodic conductance peaks as a function of $V_P$ are observed (the well-known Coulomb blockade peaks). Fig. S2.1 shows these conductance peaks in the $V_P/B$ plane. The magnetic field independent peaks fit the known behavior of an outer edge in a Coulomb dominated FPI.

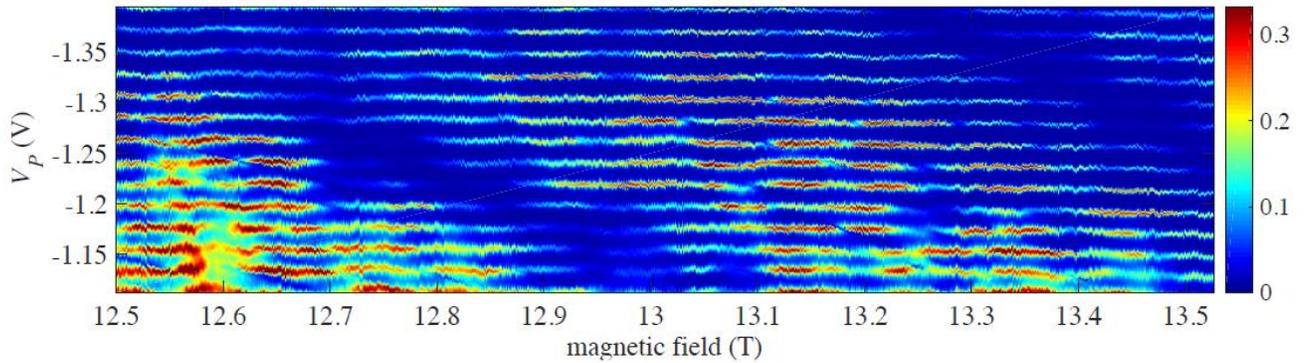

**Fig S2.1**: Conductance through the FPI as function of both $V_p$ and magnetic field. Coulomb blockade peaks without magnetic field dependence are apparent and fit the known behavior of outer edge interference in a Coulomb dominated FPI.

## S3. Temperature Dependence of the FPI peaks and dips

It is interesting to look at the temperature dependence of the conductance peaks ($t_{FPI} \ll 1$) and the conductance dips ($t_{FPI} = \frac{1}{2}$) - presented in Fig. S3.1. The conductance peaks broaden with temperature and thus overlap, increasing conductance between the peaks (Fig S3.1a). Plotting the average of the conductance peaks' height ($G_{max}$), we find an increase, nearly linear with log $T$, followed by a tendency to saturate above 100mK near $\frac{1}{3}\frac{e^2}{h}$ (Fig. S3.1b). It is expected that in a quantum dot which is weakly coupled to its leads in the fractional regime, tunneling of electrons is dominant. And thus, the temperature dependence of the conductance should follow $G_{max} \propto T^{\frac{1}{g}-2}$, with the expected Luttinger parameter $g = \frac{1}{3}$ [1]. As seen in Fig. S3.1b the actual data gives $G_{max} \propto T^{0.28 \pm 0.06}$, or $g = 0.44 \pm 0.01$; above the predicted $g = \frac{1}{3}$, but still in the in the repulsive Luttinger liquid regime ($g < 1$, [2]).

In contrast, the temperature dependence of the average of conductance dips below the $\frac{1}{3}\frac{e^2}{h}$ plateau, denoted as $\Delta G_{max}$, is very different (Figs. S3.1c). $\Delta G_{max}$ is insensitive to temperature at relatively low temperatures, $T < 80mK$, and then decreases rapidly as $\Delta G_{max} \propto T^{-3.8 \pm 0.01}$. While the decrease in $\Delta G_{max}$ coincides qualitatively with the theoretical prediction of a strongly coupled Luttinger type leads to the dot (with $e/3$ quasi-particles performing the tunneling), $\Delta G_{max} \propto T^{g-2} = T^{-1.67}$, the actual measured exponent is different.

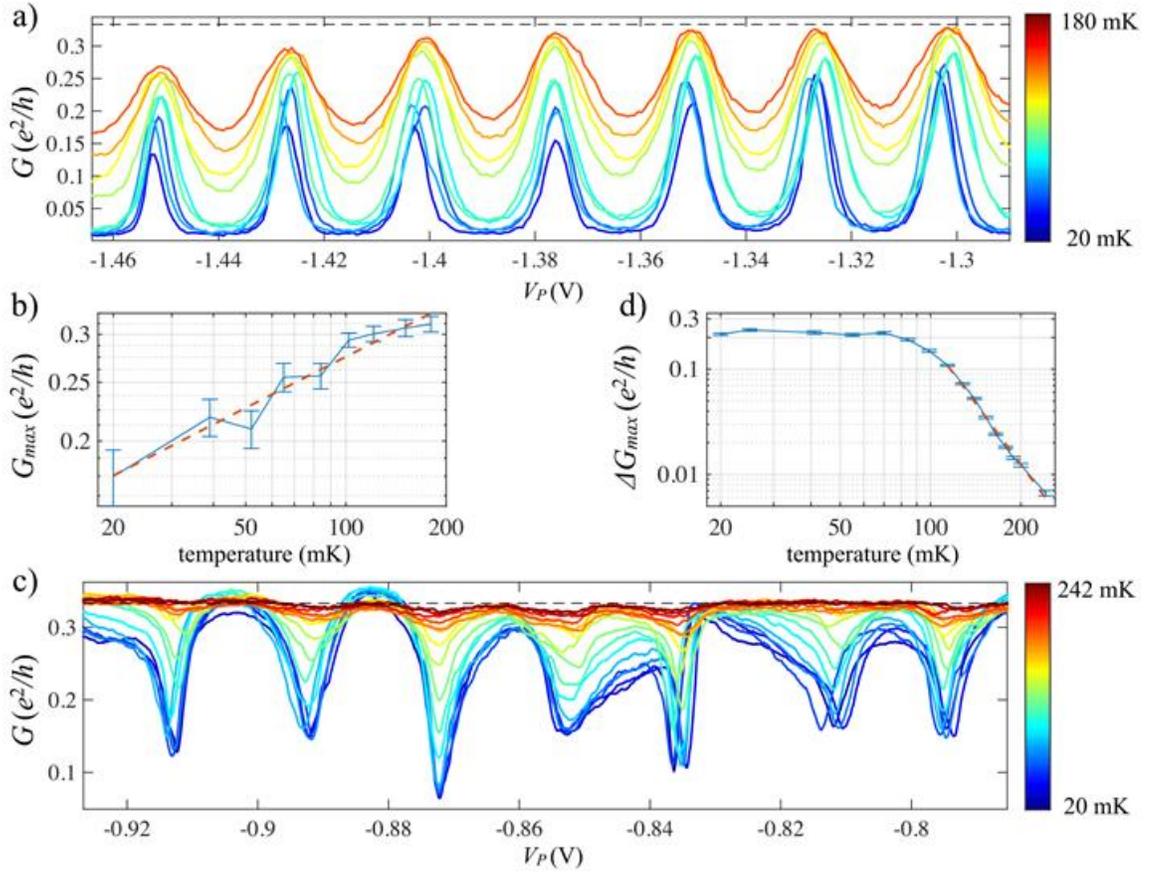

**Figure S3.1: Temperature dependence of the conductance peaks and dips:**

a) Coulomb blockade conductance peaks ($t_1, t_2 \ll 1$) at different temperatures. The black dashed line represent conductance of $\frac{1}{3}\frac{e^2}{h}$.

b) The Coulomb blockade peaks' height as function of temperature on a logarithmic scale. The fitted slope (red) gives $G \propto T^{0.37}$.

c) Conductance dips ($t_1 = t_2 = \frac{1}{2}$) below the $\frac{1}{3}\frac{e^2}{h}$ conductance plateau at different temperatures.

d) The depth of the conductance dips as function of temperature in a logarithmic scale.

## S4. FPI dips for $v = 4$

We have observed a similar effect of dips in the conductance through a FPI in the case of the integer Quantum Hall state $v = 4$, when both QPCs are set to $G = \frac{e^2}{h}$ such that one edge mode goes through the FPI and three are fully reflected and confined inside the FPI. In this setup, the conductance through the FPI shows a series of quasi-periodic dips as function of the voltage on the plunger gate. Like in the case of $v = \frac{2}{3}$, we attribute those dips to the resonant tunneling from the biased outer edge to one of the inner edges confined inside the FPI and then to the unbiased outer edge. Fig S3.1 shows an example of these dips together with regular Coulomb blockade peaks for lower $V_P$ values.

No similar effect was observed in $v = 2$ probably due to lack (or very small) tunneling between the two edge modes.

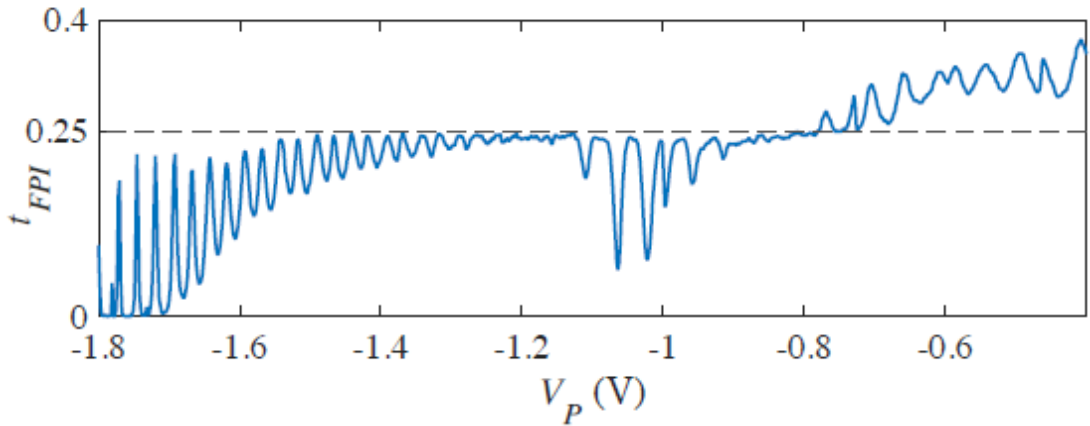

**Fig S4.1**: Transmission through the FPI as function of $V_P$ in $v = 4$ showing dips below $t = 0.25$, similar to the dips observed in $v = 2/3$. Regular Coulomb Blockade peaks are also visible.

## S5. Temperature dependence of the equilibration of the two edge modes at $\nu = \frac{2}{3}$

We have investigated the temperature dependence of the equilibration between the two $\frac{1}{3}\frac{e^2}{h}$ edge modes in $\nu = \frac{2}{3}$ using a device with a controlled separation between the two QPCs, such that the separation between them can be altered to either $L = 0.4\mu m$ or $L = 8\mu m$. In both cases we saw no temperature dependence (up to 300mK) : in the case of $L = 0.4\mu m$ no mixing between the two channels was observed, whilst in the case of $L = 8\mu m$ separation the edge was fully equilibrated to an effective single $\frac{2}{3}\frac{e^2}{h}$ charged mode for all temperatures in this range.

Fig S5.1 shows an example of these results:

- The two left sub-figures were taken from a device set to $L = 0.4\mu m$ separation between the QPCs for temperature of 25mK (upper) and 220mK(lower). In both cases $t_{S1 \to D1}$ exhibits a plateau at the value of $\frac{1}{2}$ as long $t_1 > \frac{1}{2}$, a signature of two separate charge modes.
- The two right sub-figures were taken from a device set to $L = 8\mu m$ separation between the QPCs for temperature of 25mK (upper) and 220mK(lower). In both cases $t_{S1 \to D1}$ exhibits a plateau at the value of $\frac{t_1}{2}$ (or more generally, $t_{S1 \to D1} = t_1 \times t_2$), a signature of fully equilibrated two charge modes.

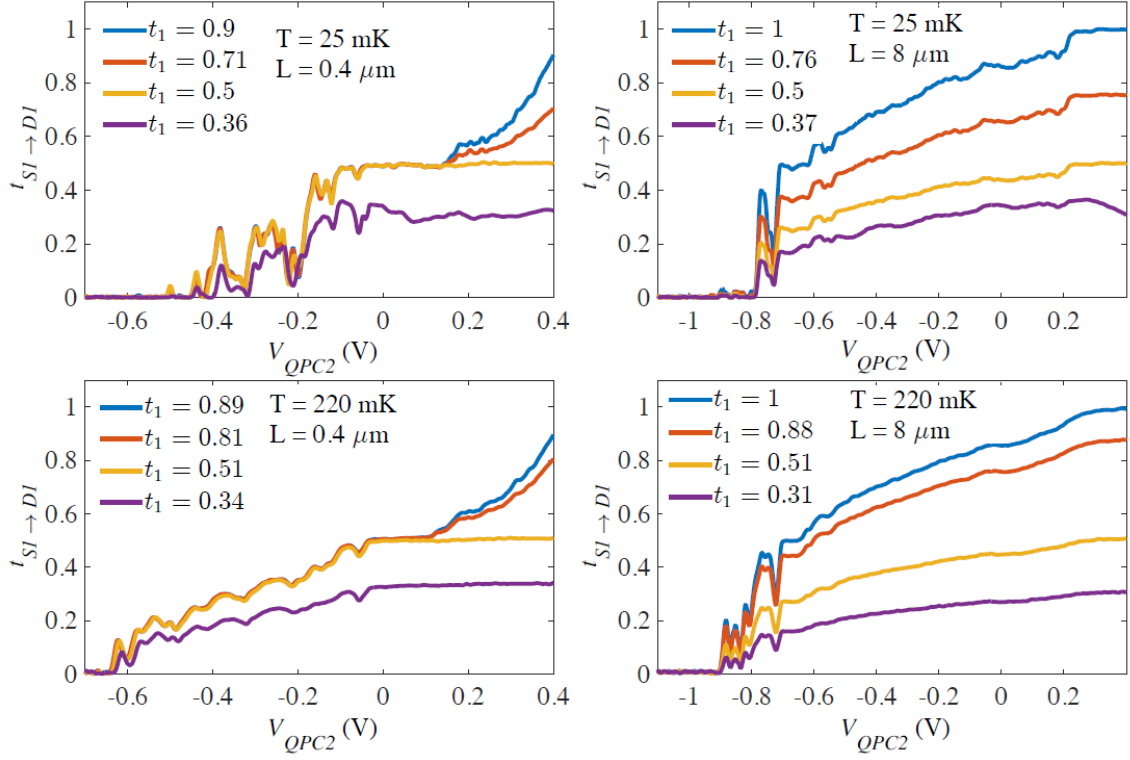

**Fig S5.1**: Set of $t_{s1\rightarrow D1}$ measurement at different QPCs separation (L) and temperatures (T) showing no temperature dependence of the edge structure.

# S6. Current Fluctuations in $\nu = \frac{3}{5}$

Figure 4a in the main text shows the measured current fluctuations in the 'two-QPC device' at $\nu = 2/3$. Here we show similar results at $\nu = 3/5$.

The current fluctuations were measured at D1 in response to current injected at S2 using the device with $L = 0.4\mu m$ separation between the QPCs (see Fig. 1b of the main text for the SEM image of the device and the definition of the different sources and drains). Setting $t_1 = 0$ the system is turned into an effective single QPC setup and exhibits a clear plateau at a value of $\frac{1}{3}\frac{e^2}{h}$ ($t = \frac{5}{9}$) as function $V_{QPC2}$. Substantial current fluctuations (shown in blue in Fig S6.1) were measured on top of this conductance plateau.

As in the case of $\nu = \frac{2}{3}$, a Fano factor cannot be assigned to these current fluctuations as no shot noise is expected to appear on top of a conductance plateau. Using the definition of 'effective Fano factor' from the main text, we get $F_{eff} = \frac{1}{3}$. Note that the effective Fano factor is defined by replacing the $t(1-t)$ term in the usual definition of Fano factor with $\frac{1}{4}$. This substitution is justified in the case of $\nu = \frac{2}{3}$ as the value of the transmission on the plateau is $\frac{1}{2}$. In the case of $\nu = \frac{3}{5}$, the transmission on the plateau is $t = \frac{5}{9}$, which yield $t(1-t) = 0.2469 \approx \frac{1}{4}$. Thus, there is no need to re-define the 'effective Fano factor' for $\nu = \frac{3}{5}$.

Setting both QPC to their conductance plateau $t_1 = t_2 = \frac{5}{9}$, the current fluctuations measured in D1 (red in Fig. S6.1) are only slightly lower than the ones measured in the 'single QPC' configuration, although no net current is reaching D1 in this case.

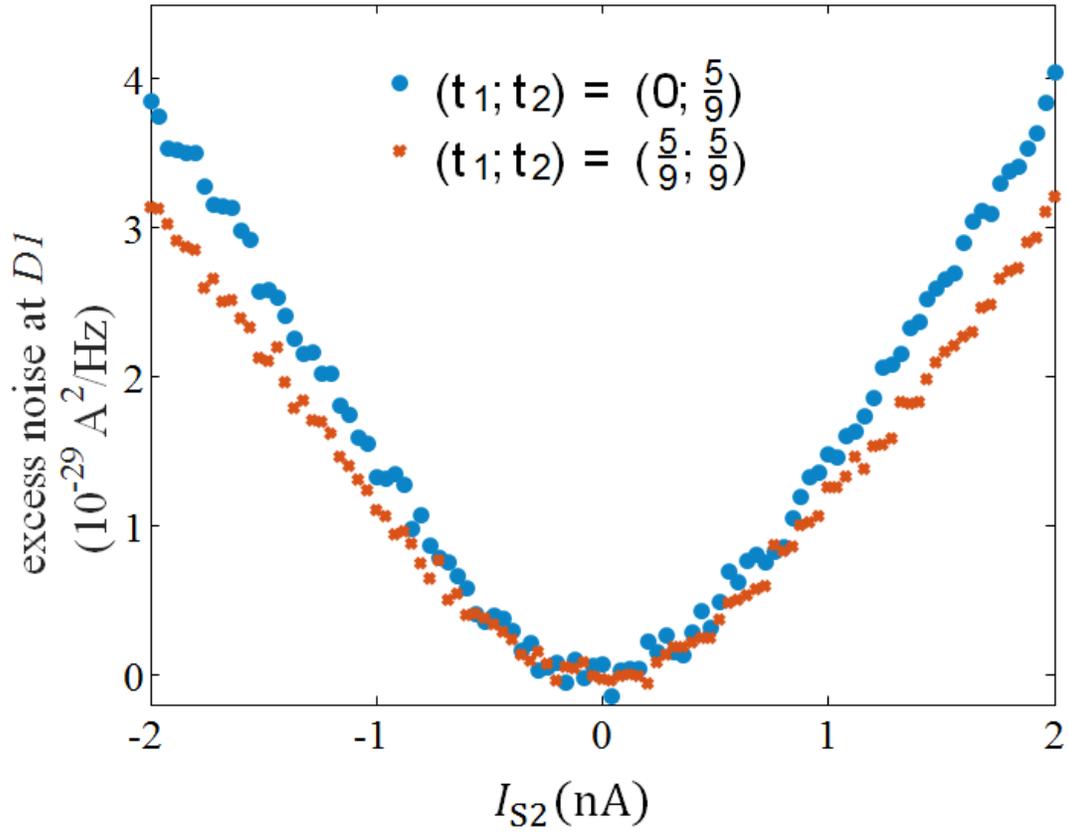

**Fig S6.1: Current fluctuations at $\nu = 3/5$.** Current fluctuations measured in D1 in response to current injected at S2 in effective single QPC configuration (blue) and the two-QPCs configuration (red).

**S7. Theoretical analysis of a two-QPC geometry**

In the main text we have discussed a theoretical model, describing a new mechanism for the generation of shot noise. This mechanism consists of a two-step process (we refer to this as a first hierarchy process): (i) charge equilibration accompanied by a generation of neutral mode excitations (neutralons or anti-neutralons, cf. Supplementary Information Section S8); (ii) the fragmentation (i.e., decay) of neutralons, leading to the stochastic creation of quasi-particle/quasi-hole pairs. This model explains well the quantized value of the Fano factor in a single QPC geometry.

Let us now apply this model to a two QPC setup. For the sake of specificity, we consider the scenario where the source S1 is biased, while the other sources are grounded [cf. Fig. 2b]. Neutral modes are excited during the equilibration processes, where a high chemical potential charge channel propagates from QPC1 (QPC2) to D3 (D1), moving in parallel to a low chemical potential channel. These neutral excitations then move toward S1 (S3), and fragment into particle-hole pairs in the charge channels propagating from S1 (S3) to D3 (D1) and from S1 (S3) to D1 (D2). Assuming that the source S1 emits 2N quasiparticles over a time interval τ; the charge $Q_{D1}$, $Q_{D2}$, and $Q_{D3}$, collected respectively at drain D1, D2, and D3, can be expressed as $Q_{D1} = \frac{e}{3}(N + \sum_{i=1}^{\frac{N}{2}} a_i - \sum_{i=1}^{\frac{N}{2}} b_i)$, $Q_{D2} = \frac{e}{3}(\sum_{i=1}^{\frac{N}{2}} b_i)$, and $Q_{D3} = \frac{e}{3}(N - \sum_{i=1}^{\frac{N}{2}} a_i)$, resulting in the effective Fano factor of 2/3, 1/3, and 1/3, respectively. Here $a_i$ ($b_i$) are random variables referring to decay of neutralons propagating from D3 (D1) to S1 (S3); they assume the value of 1 or -1, each with probability ½. The value 1 represents a creation of a quasi-particle in the outer channel and a quasi-hole in the inner channel, while -1 represents the opposite process: a creation of a quasi-hole in the outer channel and a quasi-particle in the inner one. Our experimentally found and theoretically computed values of the Fano factors for different configurations of source and drain are summarized in Fig 6. Evidently there are certain discrepancies between our theory (cf. $F_{th}^{(1)}$ in Fig 6) and the experimentally observed Fano factors (cf. $F_{exp}$ in Fig 6).

The experiment-theory agreement may be improved (cf. $F_{th}^{(2)}$ in Fig 6) when we resort to second hierarchy processes. Consider, for example, the configuration depicted in Fig 4c, featuring S1 as the source. First hierarchy processes result in a noiseless inner channel and a noisy outer channel (of the same chemical potential) entering drain D2: the outer channel flowing towards

D2 is noisy: it supports $N/2$ charge excitations, either quasi-particles or quasi-holes. By contrast, the inner channel is quiet. While the average charge carried by each of these channels is the same, this picture suggests a further equilibration process, whereby local imbalance in the charge density between the two channels gives rise to further inter-channel tunneling. Specifically, on the average $N/4$ quasi-particles or quasi-holes tunnel from the outer channel to the inner one, generating neutralons or anti-neutralons, which move from D2 to S2. These neutralons further fragment into quasi-particle/quasi-hole pairs in the charge channels propagating from S2 to D3 and from S2 to D2. There is, therefore, an additional contribution to the charge collected in the various drains, $Q_{D2} = \frac{e}{3}(\sum_{i=1}^{\frac{N}{2}} b_i - \sum_{i=1}^{\frac{N}{4}} c_i)$ and $Q_{D3} = \frac{e}{3}(N - \sum_{i=1}^{\frac{N}{2}} a_i + \sum_{i=1}^{\frac{N}{4}} c_i)$, while $Q_{D1}$ is not modified. These result in the Fano factors of 1/2 at D2 and D3. Here $c_i$ are random variables (whose values are either 1 or -1, each with probability ½), associated with the decay of neutralons propagating from D2 to S2. The value 1 represents a creation of a quasi-particle in the outer channel and a quasi-hole in the inner channel, while -1 represents the opposite process. We refer to this additional equilibration process as second hierarchy. In a single QPC, such second hierarchy processes do not take place since there are no co-propagating noisy/quiet channels (all charge channels carry the same degree of noise. As for the 'two-QPC' setup, accounting for this second hierarchy equilibration process generally improves the experiment-theory agreement (cf. column $F_{\text{th}}^{(2)}$ of the Table of Fig. 6).

## S8. Tunneling operators

Our model is based on the edge structure proposed in Ref. [3]. Moving from the edge to the bulk, each of the filling factor discontinuities, 1/3, -1/3, 1, -1/3 corresponds to a respective edge channel. The bosonic fields $\phi_j$ ($j$=1, 2, 3, 4) accounting for low energy dynamics of the corresponding channels satisfy Kac-Moody algebra $[\phi_j(x), \phi_{j'}(x')] = i\pi K_{jj'}^{-1} \text{sgn}(x - x')$; the $K$ matrix is written in a diagonal form with the elements 3, -3, 1, -3 on the diagonal. Assuming that the outermost edge channel is far from the three inner channels, one considers a two-step renormalization group procedure [S1]; in the first step only tunneling and interaction between the three inner channels is considered, leading to an intermediate fixed point with a renormalized downstream 1/3 charge channel, and two upstream neutral channels (cf. Fig. 1a, III). The respective bosonic fields, $\phi_c, \phi_{n_1}$, and $\phi_{n_2}$, may be expressed in terms of the original bosonic fields ($\phi_{j>1}$) (the outermost channel is excluded) through the matrix $M^{-1}$ ($\phi_{l=c,n_1,n_2} = (M^{-1})_{lj} \phi_{j>1}$),

$$M^{-1} = \begin{pmatrix} \sqrt{3} & \sqrt{3} & \sqrt{3} \\ \frac{3}{\sqrt{2}} & \frac{1}{\sqrt{2}} & 0 \\ \frac{1}{\sqrt{2}} & \frac{1}{\sqrt{2}} & \sqrt{2} \end{pmatrix}. \qquad (S1)$$

In this intermediate fixed point, interactions between the renormalized channels vanish.

We next note that the most general tunneling operators between the outer channel and the inner ones may be written in terms of the original bosonic modes as

$$T_{m_1,m_2,m_3,m_4} = e^{i \sum_{j=1}^{4} m_j \phi_j} = e^{i \vec{m} \cdot \vec{\phi}}, \qquad (S2)$$

with integer $\{m_j\}$. The scaling dimension of the tunneling operators is

$$2\Delta_{\vec{m}} = \frac{1}{3} m_1^2 + \left(-\frac{m_2}{\sqrt{3}} + \sqrt{3} m_3 - \frac{m_4}{\sqrt{3}}\right)^2 + \left(\frac{m_2}{\sqrt{2}} - \frac{m_3}{\sqrt{2}}\right)^2 + \left(\frac{m_2}{\sqrt{6}} - \frac{\sqrt{3} m_3}{\sqrt{2}} + \frac{\sqrt{2} m_4}{\sqrt{3}}\right)^2. \qquad (S3)$$

The most relevant quasiparticle tunneling operators transferring charge $e/3$ from the outer channel to inners are determined by two constraints. First, $e^{i m_1 \phi_1}$ is associated with annihilating a quasiparticle while $e^{i\sum_{j=2}^{4} m_j \phi_j}$ is associated with creating a quasiparticle; $m_1 = -1$ and $-\frac{m_2}{3} + m_3 - \frac{m_4}{3} = \frac{1}{3}$ should be satisfied. Second, the scaling dimension of Eq. S3 is minimized. The six most relevant quasiparticle tunneling operators from the outer channel to inners are then:

$$T_{\vec{m}^{(1)}} = e^{-i(\phi_1+\phi_2)} = e^{-i\left(\phi_1 - \frac{\phi_c}{\sqrt{3}} + \frac{\phi_{n_1}}{\sqrt{2}} + \frac{\phi_{n_2}}{\sqrt{2}}\right)},$$

$$T_{\vec{m}^{(2)}} = e^{-i(\phi_1-3\phi_2-2\phi_3-2\phi_4)} = e^{-i\left(\phi_1 - \frac{\phi_c}{\sqrt{3}} - \frac{\phi_{n_1}}{\sqrt{2}} - \frac{\phi_{n_2}}{\sqrt{2}}\right)},$$

$$T_{\vec{m}^{(3)}} = e^{-i(\phi_1-2\phi_2-\phi_3)} = e^{-i\left(\phi_1 - \frac{\phi_c}{\sqrt{3}} - \frac{\phi_{n_1}}{\sqrt{2}} + \frac{\phi_{n_2}}{\sqrt{2}}\right)},$$

$$T_{\vec{m}^{(4)}} = e^{-i(\phi_1-\phi_3-2\phi_4)} = e^{-i\left(\phi_1 - \frac{\phi_c}{\sqrt{3}} + \frac{\phi_{n_1}}{\sqrt{2}} - \frac{\phi_{n_2}}{\sqrt{2}}\right)},$$

$$T_{\vec{m}^{(5)}} = e^{-i(\phi_1-2\phi_2-2\phi_3-3\phi_4)} = e^{-i\left(\phi_1 - \frac{\phi_c}{\sqrt{3}} - \sqrt{2}\phi_{n_2}\right)},$$

$$T_{\vec{m}^{(6)}} = e^{-i(\phi_1+\phi_4)} = e^{-i\left(\phi_1 - \frac{\phi_c}{\sqrt{3}} + \sqrt{2}\phi_{n_2}\right)} \qquad (S4)$$

All of them have the same scaling dimension of $\Delta_{\vec{m}} = 2/3$, and involve the creation of neutralons. Similarly, the most relevant tunneling operators from the inners to the outer are the Hermitian conjugates of the operators in Eq. S4.

# Bibliography

[1] A. Furusaki, "Resonant tunneling through a quantum dot weakly coupled to quantum wires," *Phys. Rev. B,* vol. 57, p. 7141, 1998.

[2] M. P. A. Fisher and L. I. Glazman, "Transport in a One-Dimensional Luttinger Liquid," in *Mesoscopic Electron Transport*, vol. 345, 1997, pp. 331-373.

[3] J. Wang, Y. Meir and Y. Gefen, "Edge Reconstruction in the ν=2/3 Fractional Quantum Hall State," *Phys. Rev. Lett. ,* vol. 111, p. 246803, 2013.